\documentclass[amsmath,amsfonts,aps,nofootinbib,preprintnumbers]{revtex4}
\usepackage{graphicx}
\usepackage{amssymb}
\usepackage{epsfig}
\def\epp{{\epsilon^{\prime}}}
\begin{document}
\baselineskip=20pt
\title{Calculation of the axial charge in the $\epsilon$ and $\epsilon'$ regimes of HBChPT}
\author{Brian Smigielski}\email{smigs@u.washington.edu} \affiliation{Department of Physics,
  University of Washington\\ Box 351560, Seattle, WA 98195, USA}
\author{Joseph Wasem}\email{wasem@u.washington.edu} \affiliation{Department of Physics,
  University of Washington\\ Box 351560, Seattle, WA 98195, USA}

\preprint{NT@UW-07-09}

\begin{abstract}
The axial charge $g_{A}$ is calculated in the $\epsilon$ regime of
Heavy Baryon Chiral Perturbation Theory to order $\epsilon^3$. To
perform this calculation, we develop a technique to compute baryon
properties in the $\epsilon$ regime of Chiral Perturbation Theory.
This technique includes contributions from pion zero momentum modes
and can be used at arbitrary order, diagram by diagram, in the
$\epsilon$ regime to calculate any matrix element. Also, a
calculation of $g_{A}$ in the $\epsilon'$ regime to order
${\epsilon'}^3$ is performed. A discussion of the domain of
applicability for both the $\epsilon$ and $\epsilon'$ regimes is
also included.
\end{abstract}

\maketitle

\section{Introduction}
As computational speed has steadily advanced in recent years,
lattice QCD has been increasingly utilized as the only known method
of computing observables directly from QCD. However, even the most
advanced calculations to date use moderate lattice volumes and
unphysically large pion masses. State of the art calculations are
performed with the spatial length of the lattice being on the order
of a few Fermis, while the temporal extent is typically larger by at
most a factor of three. These calculations are also done with pion
masses that are at least a factor of two greater than the physical
pion mass. In studying low energy observables using lattice QCD it
is therefore important to fully understand how the finite volume of
the lattice and the large pion mass affect the result. To study the
low energy dynamics of QCD and finite volume effects thereof Chiral
Perturbation Theory (ChPT) in finite volume can be used.

When studying baryon properties the low energy theory used is Heavy
Baryon Chiral Perturbation Theory (HBChPT)\cite{Jenkins:1991ne}. For
infinite volume HBChPT, the small expansion parameters are
$\Lambda_{{\rm QCD}}/m_{B}$, $p/\Lambda_{\chi}$, and
$m_{\pi}/\Lambda_{\chi}$ where $p$ is the typical momentum and
$\Lambda_{\chi}$ is the chiral symmetry breaking scale typically of
order 1 GeV\cite{Luscher:1985dn}. This counting scheme also holds in
finite volume when $m_{\pi}L/2\pi>>1$\cite{Beane:2004tw}. However, a
problem exists for small quark masses ($m_{\pi}L/2\pi<<1$), as the
zero momentum mode pion propagator goes as $1/m_{\pi}^2 V$, where
$V$ is the spacetime volume\cite{Gasser:1987ah}. Thus as
computational power continues to increase and lattice calculations
are performed at ever lower quark masses, the contribution from zero
momentum mode pions will become ever larger. To account for this in
the HBChPT calculation the counting scheme must be changed to
enhance the order at which zero mode diagrams contribute.

To remedy this problem, the construction of a new power counting
scheme is needed such that the zero momentum modes are enhanced and
integrated over exactly while the non-zero momentum modes can be
treated
perturbatively\cite{Gasser:1987ah,Hansen:1990un,Hansen:1990yg,Hansen:1990kv,
Hasenfratz:1989pk,Hasenfratz:1989ux,Leutwyler:1992yt}. This leads to
the formulation of the $\epsilon$ regime. In the $\epsilon$ regime,
if $L$ and $\beta$ are, respectively, the spatial and temporal
extent of the box one is working in, $\epsilon \sim 2\pi /
\Lambda_{\chi}L \sim 2\pi / \Lambda_{\chi} \beta$ and $\epsilon^2
\sim m_{\pi} / \Lambda_{\chi}$. This counting takes into account
that the zero momentum pion contributions have become
nonperturbative. The integration over these zero modes is achieved
by utilizing the method of collective variables\cite{Gasser:1987ah}
and methods for including baryons in this framework have been
proposed previously\cite{Bedaque:2004dt}.

Another important regime for working in small volumes is the $\epp$
regime\cite{Detmold:2004ap}. This regime is characterized by a
highly asymmetric hyperbox with spatial dimension $L$ and a large
temporal dimension. Due to the zero mode pion propagator mentioned
above (which goes as $1/m_{\pi}^{2}V$), as one approaches the
$\epsilon$ regime from the $p$ regime the spatial and temporal zero
modes will become enhanced relative to the nonzero modes. While they
do not yet need to be treated nonperturbatively, their counting does
need to be enhanced relative to the nonzero modes. To account for
this, the $\epp$ regime possesses similar counting rules to the
$\epsilon$ regime: $\epp \sim 2\pi / \Lambda_{\chi}L$ and $ {\epp}^2
\sim m_{\pi} / \Lambda_{\chi}$, but separate rules for the temporal
counting. For nonzero momentum modes the time direction counts
according to $\epp \sim 2\pi / \Lambda_{\chi}\beta$. However, for
pion loops containing spatial zero modes this counting changes to
${\epp}^2 \sim 2\pi / \Lambda_{\chi}\beta$. In this regime, zero
modes remain perturbative. For the specific combinations of the
lattice size and quark mass used in this paper, the $\epsilon$ and
$\epp$ regimes are the correct counting schemes to use.

The axial-vector current has been previously studied in the $p$
regime in infinite
volume\cite{Jenkins:1991es,Zhu:2000zf,Hemmert:2003cb,Beane:2004rf,Khan:2006de,Bernard:2006te,Procura:2006gq}
as well as in finite volume\cite{Beane:2004rf} and provides an
important cornerstone of lattice QCD efforts to understand baryon
physics\cite{Capitani:1999zd,Dolgov:2002zm,
Ohta:2004mg,Khan:2004vw,Edwards:2005ym,Khan:2006de}. This work
focuses on understanding the nonperturbative aspects of the
$\epsilon$ regime and computing the axial-vector current matrix
element between two nucleon states of equal momenta for lattice
sizes that require the use of the $\epsilon$ and $\epp$ regimes of
HBChPT. By better understanding how the volume of the lattice and
choice of quark mass determines which power counting scheme is
needed, one can more accurately extrapolate the infinite volume,
physical pion mass results for $g_A$ from the lattice data.

\section{Heavy Baryon Chiral Perturbation Theory}
At low energy one can use HBChPT to describe the dynamics of
nucleons and pions. The Lagrangian that is consistent with
spontaneously broken $SU(2)_{L}\otimes SU(2)_{R}$ is at leading
order\cite{Jenkins:1990jv,Jenkins:1991ne}:
\begin{eqnarray}\label{lagrangian}
    \mathcal{L}_0&=&\bar{N}iv\cdot\mathcal{D}N-\bar{T}_{\mu}iv\cdot\mathcal{D}T^{\mu}+\Delta\bar{T}_{\mu}T^{\mu}
    +\frac{f^2}{8}{ \rm Tr}[\partial_{\mu}\Sigma^{\dagger}\partial^{\mu}\Sigma]+\lambda\frac{f^2}{4}{ \rm Tr}[m_{q}\Sigma^{\dagger}+h.c.]\nonumber\\
    &&+2g_{A}^{(0)}\bar{N}S^{\mu}\mathcal{A}_{\mu}N
    +g_{\Delta N}[\bar{T}^{abc,\nu}\mathcal{A}^{d}_{a,\nu}N_{b}\epsilon_{cd}+h.c.]+2g_{\Delta\Delta}\bar{T}_{\nu}S^{\mu}\mathcal{A}_{\mu}T^{\nu}
\end{eqnarray}
with the velocity dependent nucleon fields $N$ (for brevity in this
paper the usual subscript $v$ and the integral over the velocity
have been dropped from the nucleon fields), the Rarita-Schwinger
fields $T^{\mu}$ describing the $\Delta$-resonances, and the
definitions:
\begin{eqnarray}
    \Sigma&=&\xi^{2}={\rm exp}\left(\frac{2iM}{f}\right),\nonumber\\
    M&=&\left(\begin{matrix}
    \pi^0/\sqrt{2} & \pi^+\cr \pi^- &
    -\pi^0/\sqrt{2}
    \end{matrix}\right), \nonumber\\
    \mathcal{A}^{\mu}&=&\frac{i}{2}(\xi\partial^{\mu}\xi^{\dagger}-\xi^{\dagger}\partial^{\mu}\xi),\nonumber\\
    V^{\mu}&=&\frac{1}{2}(\xi\partial^{\mu}\xi^{\dagger}+\xi^{\dagger}\partial^{\mu}\xi),\nonumber\\
    \mathcal{D}^{\mu}&=&\partial^{\mu}+V^{\mu}.
\end{eqnarray}
The pion fields are encapsulated in the matrix $M$. The
Rarita-Schwinger fields are rank-3 tensors such that:
\begin{equation}
    \begin{matrix}
    T^{111}=\Delta^{++}, & T^{112}=\frac{1}{\sqrt{3}}\Delta^{+}, & T^{122}=\frac{1}{\sqrt{3}}\Delta^{0},
    & T^{222}=\Delta^{-}
    \end{matrix}
\end{equation}
while the nucleons are simply an SU(2) vector given by
\begin{equation}
    N=\left(\begin{matrix}
    p\cr n \end{matrix}\right)
\end{equation}
where the constant $f=132$ MeV and the matrix $m_q$ is the quark
mass matrix. The three couplings given in eqn. (\ref{lagrangian})
($g_{A}^{(0)}$, $g_{\Delta N}$, and $g_{\Delta\Delta}$) are the
infinite volume, chiral limit couplings between baryons and pions.
Also, $S^{\mu}$ is the covariant spin vector and $v^{\mu}$ is the
heavy baryon four velocity with $v^2=1$ (typically
$v^{\mu}=(1,\vec{0})$).

In our calculation higher order Lagrangian terms will also become
important. The relevant next to leading order Lagrangian terms are:
\begin{eqnarray}\label{1/m lagrangian}
    \mathcal{L}_{1}&=&-\left(\bar{N}\frac{\mathcal{D}^2-(v\cdot\mathcal{D})^2}{2m_B}N\right)
    +\left(\bar{T}^{\mu}\frac{\mathcal{D}^2-(v\cdot\mathcal{D})^2}{2m_B}T_{\mu}\right)\nonumber\\
    &&+\lambda C_{1}\bar{N}{ \rm Tr}[m_{q}\Sigma^{\dagger}+h.c.]N
    +4\bar{N}\left(C_{2}-\frac{{g_{A}^{(0)}}^{2}}{8m_B}\right)\mathcal{A}_{0}^{2}N+4C_{3}\bar{N}\mathcal{A}^{\mu}\mathcal{A}_{\mu}N \nonumber \\
    &\longrightarrow&\bar{N}\frac{\vec{\partial}^2}{2m_B}N-\bar{T}^{\mu}\frac{\vec{\partial}^2}{2m_B}T_{\mu}
    +\lambda C_{1}\bar{N}{ \rm Tr}[m_{q}\Sigma^{\dagger}+h.c.]N
    +4\bar{N}\left(C_{2}-\frac{{g_{A}^{(0)}}^{2}}{8m_B}\right)\mathcal{A}_{0}^{2}N+4C_{3}\bar{N}\mathcal{A}^{\mu}\mathcal{A}_{\mu}N
\end{eqnarray}
where the chirally covariant derivatives $\mathcal{D}$ are converted
to normal derivatives $\partial$ as the contributions from $V^{\mu}$
will not be important at the order considered. The coefficients in
front of these terms can be determined by reparametrization
invariance\cite{Luke:1992cs} or by matching to the relativistic
theory in the path integral\cite{Bernard:1992qa}. The last three
terms are not important for calculating the axial matrix element to
order $\epsilon^3$ or $\epp^3$, but are important for the nucleon
mass. The first two terms in eqn. (\ref{1/m lagrangian}) are simple
higher order extensions of the kinetic energy operators given in the
lowest order Lagrangian. Including these terms modifies our baryon
and decuplet propagators. It should also be noted that these two
order $1/m_B$ terms are not the only $1/m_B$ operators, but they are
the only ones that will enter in the evaluation of the axial matrix
element to the order we work in $\epsilon$ and $\epp$. Defining
$\tau^a_{\xi_{+}}=\frac{1}{2}(\xi^{\dagger}\tau^{a}\xi+\xi
\tau^{a}\xi^{\dagger})$ and
$\tau^a_{\xi_{-}}=\frac{1}{2}(\xi^{\dagger}\tau^{a}\xi-\xi
\tau^{a}\xi^{\dagger})$, the axial-vector current is:
\begin{eqnarray}
    j_{\mu 5}^{a}&=&g_{A}^{(0)}\bar{N}S^{\mu}\tau^a_{\xi_{+}}N + \frac{1}{2}g_{\Delta
    N}\left(\bar{T}^{abc,\mu}(\tau^a_{\xi_{+}})^{d}_{a}N_{b}\epsilon_{cd}+h.c.\right) +
    g_{\Delta \Delta}\bar{T}S^{\mu}\tau^a_{\xi_{+}}T+\bar{N}v^{\mu}\tau^a_{\xi_{-}}N+\dots
\end{eqnarray}
where the ellipses indicate higher order terms. The meson part of
the current has been omitted as it is not necessary for what
follows. The form of the higher order current is easily derivable by
beginning with the fully relativistic expression for the current,
and then substituting in the appropriate expansion for the
relativistic field in terms of the heavy field to the desired order
in $1/m_{B}$. Upon doing this, and working to
$\mathcal{O}(\frac{1}{m_{B}})$, the higher order current
is\cite{Falk:1990pz,Walker-Loud:2004hf}:
\begin{eqnarray}
j_{\mu 5}^{a,(1)}&=&\bar{N}\Gamma_{\mu}^{a}\left(\frac{i
\roarrow{D}\!\!\!\!/_{\perp}}{2m_{B}} \right)N -
\bar{N}\left(\frac{i \loarrow{D}\!\!\!\!/_{\perp}}{2m_{B}}
\right)\Gamma_{\mu}^{a}N
\end{eqnarray}
where $D\!\!\!\!/_{\perp}=D\!\!\!\!/-v\!\!\!/ \ (v \cdot D)$,
$\Gamma_{\mu}^{a}=S_{\mu}\tau^a_{\xi_{+}}$, and the superscript
$(1)$ denotes $1/m_B$ suppressed contribution to the current.
Working to lowest order in the above current amounts to the
replacement of $D\!\!\!\!/$ by $\partial\!\!\!/$. At tree level,
this contribution vanishes as it is proportional to the spatial
on-shell momenta which is taken to zero at the end. Similarly,
including this operator within a loop diagram yields zero. This can
be understood simply since both the pion propagator and the two
pion-baryon vertices are even with respect to the spatial loop
momenta whereas the higher order current supplies an odd power of
the spatial loop momenta. These higher order terms will not be
necessary for this calculation.

In finite volume, the fully relativistic nucleon field must satisfy
the conditions\cite{Bedaque:2004dt}:
\begin{eqnarray}
    \psi(t,\vec{r}+\vec{n}L)&=&-\psi(t,\vec{r}), \nonumber \\
    \psi(\beta,\vec{r})&=&-\psi(0,\vec{r})
\end{eqnarray}
where $\beta$ is the temporal extent of the volume. Choosing the
rest frame of the nucleon $v^{\mu}=(1,0,0,0)$, this then implies that the heavy nucleon
field satisfies the condition:
\begin{equation}
    N(\beta,\vec{r})=-e^{\beta m_B}N(0,\vec{r}).
\end{equation}
This means that the heavy nucleon field has the Fourier decomposition
\begin{equation}\label{baryonfourier}
    N(\tau,\vec{r})=\sum_{n_0}{\rm
    exp}\left[-i\left(\frac{\pi(2n_0+1)}{\beta}+im_B\right)\tau\right]N(0,\vec{r}).
\end{equation}
A similar condition exists for the decuplet fields. The presence of the baryon mass in the above equation will become important when looking at the finite time corrections to
Feynman graphs.

\section{$\epsilon$ and $\epp$ Expansion Regimes of Validity}
\begin{figure}[!h]
\centering
  \includegraphics[scale=.75]{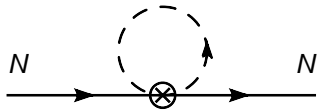}
  \caption{One Loop Contribution}\label{regimeloop}
\end{figure}

For a nucleon in a finite volume there are four different regimes
that are dependent on the choices of the quark mass (and by
extension the pion mass), the spatial extent of the volume, and the
temporal extent of the volume. These regimes define particular power
countings appropriate for the physics and are variously called
$\epsilon$, $\delta$, and $\epp$\cite{Gasser:1987ah,
Leutwyler:1987ak, Detmold:2004ap,Luscher:1985dn} regimes, along with
the standard $p$ regime. A simple illustration of the boundaries
between them can be provided by examining  how the different modes
contained in the simple pion loop in fig. (\ref{regimeloop}) count.
This definition of the boundaries is not unique but serves as a
useful guide.
Concentrating on the $\epsilon$, $\delta$, and $\epp$ regimes, each
of these three expansion parameters are similarly defined by the
relation:
\begin{equation}
    \epsilon\sim{\epp}\sim\delta\sim\frac{2\pi}{\Lambda_{\chi}L}.
\end{equation}
A useful relation to separate these regimes from the $p$ regime is
the condition
\begin{equation}\label{pregimebound}
    \frac{m_{\pi}L}{2\pi}<1
\end{equation}
which is a quantity that counts as $\epsilon$ or $\epp$, and allows
for a perturbative expansion in $m_{\pi}L$.

Examining the different modes of fig. (\ref{regimeloop}) yields:
\begin{eqnarray}
    I&\sim&\frac{1}{\beta L^3}\sum_{n_\mu}\frac{1}{q_{0}^{2}+|\vec{q}|^{2}+m_{\pi}^{2}}
    \sim\left[\begin{matrix}\frac{1}{\beta L^{3} m_{\pi}^{2}}, & \ \  q_{\mu}=(0,\vec{0}) \\
    \cr \frac{\beta}{L^3}, & \ \ q_{\mu}=(q_0,\vec{0}) \\
    \cr \frac{1}{\beta L}, & \ \ q_{\mu}=(q_0,\vec{q}) \end{matrix}\right]
\end{eqnarray}
where in finite volume one has the definitions $|\vec{q}|^2=(2\pi)^2
(n_1^2+n_2^2+n_3^2)/L^2$ and $q_0^2=(2\pi)^2 n_0^2/\beta^2$. The
spatial and temporal zero modes are then defined by $\vec{n}=0$ and
$n_0=0$ respectively. To separate the three regimes ($\epsilon$,
$\delta$, and $\epp$) one must look at the boundaries of where both
the spatial and the temporal zero mode enter at order one.

Also of interest is the quantity $\Delta L/2\pi$. It should be noted
that strictly speaking $\Delta L >> 1$ because $\Delta \sim
\Lambda_{{\rm QCD}} \sim f_{\pi}$ and in order for hadronic physics
to be contained within the volume we work, $f_{\pi}L>>1$.
Calculationally however, $\frac{\Delta L}{2\pi}$ is numerically the
same as $\frac{m_{\pi}L}{2\pi}$ in the region of the $\epsilon$ and
$\epp$ regimes that are of primary interest to lattice QCD
calculations, and this allows for a perturbative expansion of
$\Delta L/2\pi$.

\begin{figure}[!h]
\centering
\includegraphics[scale=.65]{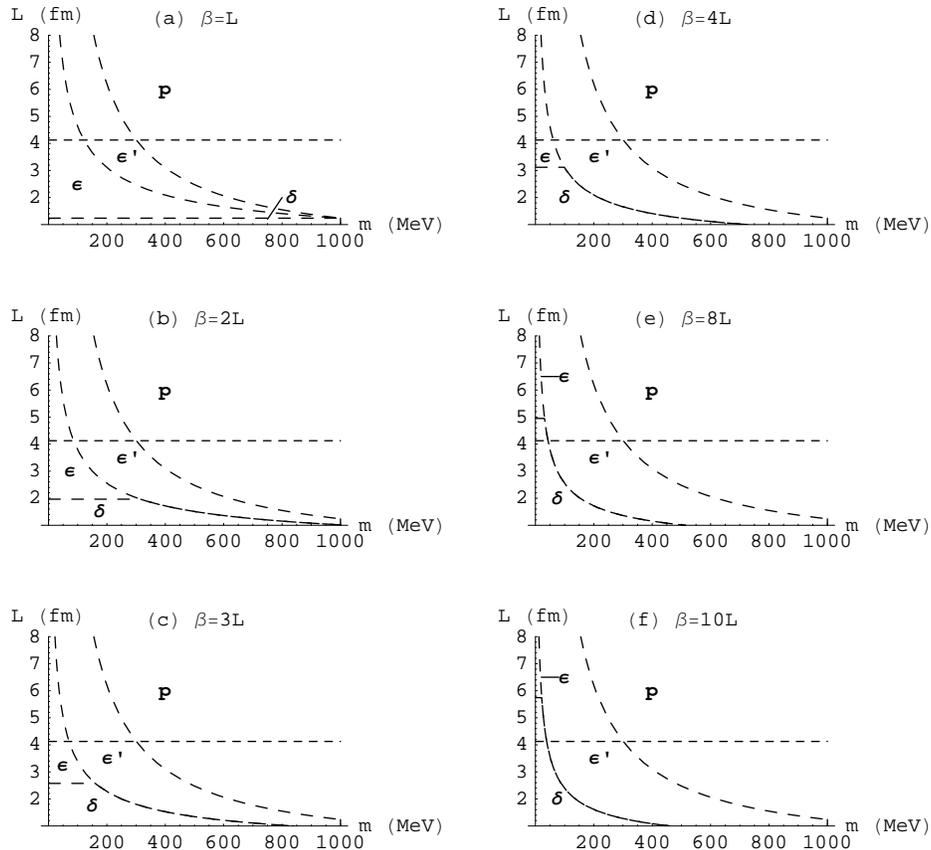}
\caption{Different counting regimes in $L$-$m_\pi$ space. The
horizontal dashed line gives the value of $2\pi/\Delta$ above which
the expansion taken in $\Delta L$ is no longer valid.
The dashed lines demarcating the boundaries between the
different regions are there to remind the reader that these are
poorly defined boundaries.} \label{etaregimes}
\end{figure}

In the $\epsilon$ regime it is required that the zero mode
$q_{\mu}=(0,\vec{0})$ counts as order one, while the $\delta$ regime
also requires that the spatial zero modes $q_{\mu}=(q_0,\vec{0})$
count as order one. In the $\epp$ regime all zero mode contributions
are perturbative. The boundaries of the three regimes in $L$-$m_\pi$
space for different values of the ratio $\beta/L$ are plotted in
fig. (\ref{etaregimes}). However, it must be emphasized that these
boundaries are necessarily very poorly defined in their placement,
and should be interpreted more as midpoints in the smooth transition
between regimes than as hard lines demarcating each region.

Within each of the regimes, how the pion mass, the decuplet mass
splitting, and zero mode $\beta$ count will vary depending on how
the ratios $m_{\pi}/\Lambda_\chi$, $\Delta/\Lambda_\chi$, and
$2\pi/\Lambda_{\chi}\beta$ compare to the ratio
$2\pi/\Lambda_{\chi}L$. This will in turn affect the order at which
certain graphs contribute. In this work
$m_{\pi}/\Lambda_{\chi}\sim\epsilon^2\sim{\epp}^2$ and
$\Delta/\Lambda_{\chi}\sim\epsilon^2\sim{\epp}^2$. However, it
should be noted that as one moves toward lattices with smaller
spatial dimension and smaller pion mass both $m_\pi/\Lambda_{\chi}$
and $\Delta/\Lambda_{\chi}$ will numerically become smaller relative
to $2\pi/\Lambda_{\chi}L$ and so will count at higher order. This
will require a reassessment of the order at which the graphs
contribute.

Additionally, the order at which $m_\pi$ contributes will have
nontrivial consequences for how $1/\beta$ is counted within the
$\epp$ regime. In this work the zero mode counting of $1/\beta$ in
the $\epp$ regime is determined by $m_\pi$, as this is where the
zero mode of the tadpole diagram of fig. (\ref{regimeloop}) becomes
large\cite{Detmold:2004ap}. However, in assigning $1/\beta\sim
m_\pi$ the tadpole zero mode counts not as $1/\beta L^3 m_{\pi}^2$
but rather as $1/L^3 m_{\pi}$. This difference determines a boundary
within the $\epp$ regime above which the counting of $1/\beta\sim
m_\pi$ for a zero mode diagram is legitimate, and below which it is
not. This line does not demarcate the boundary between $\epsilon$
and $\epp$, but it does determine the regions where $1/ \beta \sim
m_{\pi}$ counting is appropriate.

\section{Evaluation of the Zero Modes for the $\epsilon$ regime}
For the $\epsilon$ regime, the zero modes, $q_{\mu}=(0,\vec{0})$,
must be treated nonperturbatively due to the $1/m_{\pi}^2 V$
dependance of the pion zero mode propagator. Starting with a purely
mesonic theory the Euclidean partition function has the form
\begin{equation}
    Z=\int\left[\mathcal{D}\Sigma\right]{\rm exp}\left[-\int d^4x \
    \mathcal{L}(\Sigma(x))\right].
\end{equation}
The group element $\Sigma(x)$ parameterizes the meson fields. The
zero and non-zero modes are described by the Fourier components of
the pion fields, $q_{n}^a$, in finite volume:
\begin{equation}
\Sigma(x)={\rm exp}\left(\frac{2i M(x) \cdot \tau}{f}\right), \ \
M^a(x)=\sum_{n}q_n^a u_n(x),
\end{equation}
where $n$ is a four-vector with integer entries, $\tau^a$ is one-half
the Pauli matrices $\sigma^a$, and the $u_n(x)$'s
are plane waves. Following the procedure of ref.
\cite{Gasser:1987ah}, a change of variables is made such that
the zero modes are separated from the non-zero modes:
\begin{eqnarray}\label{chgofvar}
\Sigma &\to& U\hat{\Sigma}U, \label{sigmachgvars} \\
U&=&{\rm exp}\left(\frac{i \phi \cdot \tau}{f}  \right), \\
\hat{\Sigma}(x)&=&{\rm exp}\left(\frac{2i \hat{M}(x) \cdot \tau}{f}\right), \\
\hat{M}^a(x)&=&\sum_{n_{\mu} \neq 0}p_{n_{\mu}}^a u_{n_{\mu}}(x).
\end{eqnarray}
The $U$'s are spacetime independent SU(2) matrices containing the
zero mode contributions while $\hat{\Sigma}(x)$ contains the non-zero
modes with new Fourier coefficients $p_n^a$. Note the $p_n^a$'s are related non-linearly to
the $q_n^a$'s. Using the Baker-Campbell-Hausdorff formula, all the
exponentials in eqn (\ref{sigmachgvars}) can be combined. The resulting exponential has the form:
\begin{eqnarray}
U\hat{\Sigma}U&=&{\rm exp}\left(\frac{2i}{f} \phi^a\tau^a+\frac{2i}{f} \hat{M}^b(x)\tau^b+\frac{1}{2f^2}[2i \phi^a\tau^a , 2i \hat{M}^b(x)\tau^b]+\dots\right) \nonumber \\
&=&{\rm exp}\left(\frac{2i}{f} \phi^a\tau^a+\frac{2i}{f} \sum_{n \neq 0}p_n^b u_n(x)\tau^b+\frac{1}{2f^2}\sum_{n \neq 0}u_n(x)[2i \phi^a\tau^a , 2i p_n^b \tau^b]+\dots\right)
\end{eqnarray}
where the ellipsis represents an infinite string of commutators
involving the $\phi$ and $\hat{M}$ matrix. The commutators will
always contain at least one power of $\hat{M}$ making them
$\mathcal{O}(\epsilon)$ or higher. Each of these matrices can in
turn be written as sums of their Fourier coefficients multiplied by
their appropriate plane wave functions as done above. Since plane
waves possess the property that $u_{n_1}(x)u_{n_2}(x) \dots
u_{n_m}(x)=u_{n_1+n_2+\dots+n_m}(x)$ one can expand all the sums and
gather all those terms which are multiplied by
$u_{n_1+n_2+\dots+n_m}(x)=u_{n_k}(x)$ for specific $n_1,n_2,\dots$
and redefine whatever function of the $p_n^a$'s and $\phi^a$'s that
correspond to it, as $q_{n_k}^a$ for $n_k \neq 0$.

Given the change of variables in eqn. (\ref{chgofvar}), one must
shift the measure of integration from $\mathcal{D}\Sigma$ to
$\mathcal{D}\hat{\Sigma}$ and $\mathcal{D}U^2$. In doing so, one
needs to calculate a Jacobian factor that is derived from a
Fadeev-Popov procedure that is utilized in ref. \cite{Hansen:1990un}
or equivalently from a determinant of the metric of the manifold the
fields live on as in ref. \cite{Gasser:1987ah}. After performing the
field redefinition, this factor can be expanded in powers of the
non-zero modes of the pion fields:
\begin{equation}
\left[\mathcal{D}\Sigma\right]
=\left[\mathcal{D}U^2\right]\left[\mathcal{D}\hat{\Sigma}\right]
\left(1+\mathcal{O}(\hat{M}^2)\right).
\end{equation}
For our calculations only the leading order piece is needed. When
including nucleons the $\xi(x)$ field is used to describe the mesons
in addition to $\Sigma(x)$, such that $\xi^2=\Sigma$. Under SU(2)
left and right transformations, $\Sigma \to L\Sigma R^{\dagger}
\implies \xi \to L\xi V^{\dagger}$ or $\xi \to V \xi R^{\dagger}$
while the nucleons transform according to $N \to VN$ such that $V$
is an SU(2) matrix that is a function of $L, R$, $\xi$, and $x$.
Under SU(2) vector transformations, $V$ reduces to a spacetime
independent element of ${\rm SU(2)}_{V}$ to ensure the correct
transformation property of the nucleons. $V$ is defined implicitly
through the following relations:
\begin{eqnarray}
V&=&\sqrt{L\Sigma R^{\dagger}} \ R \ \xi^{\dagger} \\
V^{\dagger}&=&\xi^{\dagger} \ L^{\dagger} \ \sqrt{L\Sigma
R^{\dagger}}
\end{eqnarray}
In building chirally invariant terms for the Lagrangian one must use
both transformation rules as stated above. It is for this
reason that the change of variables needed to separate the zero
modes from $\xi$ are given by the following:
\begin{eqnarray}
\xi &\to& U\hat{\xi}V^{\dagger} \label{xichgvars1} \\
\xi &\to& V \hat{\xi} U \label{xichgvars2}
\end{eqnarray}
and therefore $V^{\dagger}=\hat{\xi}^{\dagger} \ U^{\dagger}
\sqrt{U\hat{\Sigma}U} $ and $V=\sqrt{U\hat{\Sigma}U} \ U^{\dagger} \hat{\xi}^{\dagger}$
where $L=R^{\dagger}=U$.

In order to compute the square root of a matrix,
assume that there exists a matrix $D$ such that
$D^2=U\hat{\Sigma}U$. Since the non-zero modes are suppressed by a
power of $\epsilon$, the form of $D$ is:
\begin{eqnarray}
D&=&U+\frac{i \epsilon}{f}A-\frac{\epsilon^2}{2f^2}B+ \dots \label{sqrtsigma}
\end{eqnarray}
where the factors of $\epsilon$ are made explicit, factors of $i/f$ were put in
for later convenience, and $A$ and $B$ can in general depend on the $p_n^a$'s and $\phi^a$. One can
expand $U\hat{\Sigma}U$ order by order in $\epsilon$ and set up a
matrix equation to determine $A$ and $B$:
\begin{eqnarray}
D^2&=&U^2+\frac{i \epsilon}{f}(UA+AU)-\frac{\epsilon^2}{2f^2}\left(UB+BU+2A^2\right)+\dots \\
U\hat{\Sigma}U&=&U^2+\frac{2i\epsilon}{f}U\hat{M}^a\tau^aU-\frac{2\epsilon^2}{f^2}U(\hat{M}^a\tau^a)^2U+\dots
\end{eqnarray}

Hence, one is able to
construct $D$ to the desired order and $V$ can be readily obtained.
Using this to calculate $V$ one obtains
\begin{eqnarray}\label{V}
V&=&\mathbf{1}+\frac{i\epsilon}{f}(-\hat{M}+AU^{\dagger})-\frac{\epsilon^2}{2f^2}\left(\hat{M}^2
-2AU^{\dagger}\hat{M}+BU^{\dagger}
\right)+\mathcal{O}(\epsilon^3) \\
\hat{M}&=&\left(\begin{matrix}
    \hat{\pi}^0/\sqrt{2} & \hat{\pi}^+\cr \hat{\pi}^- &
    -\hat{\pi}^0/\sqrt{2}
    \end{matrix}\right) \label{Mhat}
\end{eqnarray}
In general $A$ and $B$ are complicated functions of the parameters
of $U$ and $\hat{M}$. The form of the matrix $A$ is provided in
appendix D, while the matrix $B$ is not necessary for the
calculation at the order considered.

Under the field redefinition using eqns. (\ref{sigmachgvars},
\ref{xichgvars1}, \ref{xichgvars2}) a few terms in the Lagrangian
will be examined to see how they are altered. For the pion-nucleon
coupling term given by
$g_{A}^{(0)}N^{\dagger}S_{\mu}\mathcal{A}^{\mu}N$, the redefinition
causes $\mathcal{A}_{\mu} \to V\hat{\mathcal{A}_{\mu}}V^{\dagger}$,
and so expanding the first couple of terms:
\begin{eqnarray}
g_{A}^{(0)}N^{\dagger}S_{\mu}\mathcal{A}^{\mu}N \to g_{A}^{(0)}N^{\dagger}S_{\mu}\left(-\frac{2i}{f}\partial^{\mu}
\hat{M}+\frac{2}{f^2}\left\lbrace [\partial^{\mu}
\hat{M},\hat{M}]+AU^{\dagger}\partial^{\mu}\hat{M}-\partial^{\mu}\hat{M}UA^{\dagger}
\right\rbrace +\mathcal{O}\left(\epsilon^3\right) \right)N \label{onepionpiece}
\end{eqnarray}
It is alarming that $\mathcal{A}_{\mu}$ now appears to contain terms which are
even in the non-zero mode pion fields since this would violate
parity. Generically denoting a zero mode pion field as $\pi^z$,
then by comparing the $U$ matrix in the form of:
\begin{equation}
U={\rm exp} \ \left[\frac{i}{f}\left(\begin{matrix}
    \pi^z_0/\sqrt{2} & \pi^z_{+}\cr \pi^z_{-} &
    -\pi^z_0/\sqrt{2}
    \end{matrix}\right)\right]
\end{equation}
with that when it is parameterized using hyperspherical
coordinates\footnote{A precise definition of the coordinates used
can be found in appendix C. Normally there is a radial coordinate
which we denote by $|b|$ that must exist to parameterize in terms of
hyperspherical coordinates. However as shown in the appendix, this
factor will be sent to one by the presence of a delta function so we
explicitly do not write it for this reason.}:
\begin{eqnarray}\label{hypersphere}
    U=\mathbf{1}{\rm cos}(\psi)+i\sigma_{1}{\rm sin}(\psi){\rm sin}(\theta){\rm cos}(\phi)+i\sigma_{2}{\rm sin}(\psi){\rm sin}(\theta){\rm sin}(\phi)
    +i\sigma_{3}{\rm sin}(\psi){\rm cos}(\theta)
\end{eqnarray}
leads to the relations:
\begin{eqnarray}
\frac{\pi_0^{z}}{f}&=&\sqrt{2}\psi \ {\rm cos}(\theta) \nonumber \\
\frac{\pi_{+}^{z}}{f}&=&\psi \ {\rm sin}(\theta) \ {\rm exp}(-i\phi) \nonumber \\
\frac{\pi_{-}^{z}}{f}&=&\psi \ {\rm sin}(\theta) \ {\rm exp}(i\phi).
\nonumber
\end{eqnarray}
Substituting eqns. (\ref{hypersphere}, \ref{Mhat}) in eqn.
(\ref{onepionpiece}), Taylor expanding the functional dependence in
$\psi$, and then using the relationships above yield terms with an
odd number of non-zero and zero mode pions multiplied by powers of
$\psi^2$. Because
$\psi^2=(\frac{1}{2}(\pi_0^z)^2+\pi_{+}^z\pi_{-}^z)/f^2$, this term
will always be even under parity. Hence, the entire contribution is
therefore explicitly even under parity. This has been checked to
$\mathcal{O}(1/f^3)$ and will hold for all orders in $1/f$.

The relationship between what are called the non-zero modes under
the field redefinition $(\Sigma \to U\hat{\Sigma}U)$ versus the
non-zero modes in $\Sigma$ must also be carefully explained. For
clarity, an expansion of $\mathcal{A}_{\mu}$ to three pion fields
with exactly one nonzero mode pion will be considered. In the
expansion of the pion-nucleon Lagrangian term in eqn.
(\ref{onepionpiece}) the $\mathcal{O}(\epsilon)$ term seems to be
devoid of zero mode information. Indeed, from the arguments just
given, it is reasonable to assume that there should exist some
function of the zero modes which accompanies it. However, if we
again use the Baker-Campbell-Hausdorff formula to relate the
original parametrization to the one in terms of the field
redefinition one finds that (where $\hat{M}'$ encapsulates the
non-zero modes before the change of variables and $\phi$
encapsulates the zero modes):
\begin{equation}
    \hat{M}=\hat{M}'+\frac{1}{6f^2}\left[\phi,\left[\phi,\hat{M}'\right]\right]+...
\end{equation}
where $\hat{M}$ is defined by $\hat{\Sigma}={\rm
exp}\left(2i\hat{M}/f\right)$. The definition of $\hat{M}$ will
contain an infinite number of commutators, all of which will be
suppressed by a factor of $\epsilon$ or more. With this definition
one can find
\begin{eqnarray}
    \mathcal{A}_{\mu}&=&\frac{1}{f}\partial_{\mu}\hat{M}+\mathcal{O}(\hat{M}^2)\nonumber\\
    &\to&\frac{1}{f}\partial_{\mu}\left(\hat{M}'+\frac{1}{6f^2}\left[\phi,\left[\phi,\hat{M}'\right]\right]\right)
    +\dots\nonumber\\
    &=&\frac{1}{f}\partial_{\mu}\hat{M}'+\frac{1}{6f^3}\left(\phi^{2}\partial_{\mu}\hat{M}'
    -2\phi\partial_{\mu}\hat{M}'\phi+\partial_{\mu}\hat{M}'\phi^{2}\right)+\dots.
\end{eqnarray}
This result is identical to the expansion of $\mathcal{A}_{\mu}$ prior to
making the change of variables. In making the change of variables
all that happened was that some of the zero mode information was
absorbed into the definition of $\hat{M}$. Therefore it is more
appropriate to view $\hat{M}$ not as strictly the non-zero modes
of the pions but as an $\mathcal{O}(\epsilon)$ suppressed quantity with
which one can utilize perturbation theory.

For the mass term of the pions, the quark masses will be taken to be
equal, hence the change for the mass term is:
\begin{eqnarray}
\lambda \frac{f^2}{2} V m_q{\rm Tr}(\xi^2 + (\xi^{\dagger})^2) &\to& \lambda \frac{f^2}{2} V m_q{\rm Tr}(U^2 \hat{\xi}^2 + (U^{\dagger})^2(\hat{\xi}^{\dagger})^2) \nonumber \\
&\simeq& \lambda \frac{f^2}{2}V m_q{\rm Tr}(U^2 + (U^{\dagger})^2)
+\mathcal{O}\left(\epsilon^2\right). \label{massterm}
\end{eqnarray}
The first term on the RHS of eqn. (\ref{massterm}) counts as an
order one term and cannot be expanded. It will therefore have the
interpretation of a probabilistic weight function when considering
group integrals over $U^2$.

Keeping only those terms of the axial-vector current which will
contribute to the order at which we work, under the redefinition,
it goes to:
\begin{equation}
    j_{5\mu}^a=g_{A}^{(0)} \bar{N}S_{\mu}V\left(U \tau^a U^{\dagger} + U^{\dagger} \tau^a
    U\right)V^{\dagger}N.
\end{equation}

One other equivalent method of separating the zero modes circumvents
the use of the $V$ matrix entirely. With this method one simply
determines the form of $\xi=\sqrt{U\hat{\Sigma}U}$ using the same
method outlined above and uses this directly in the Lagrangian. It
has not been utilized in this paper because the redefinition in
terms of the $V$'s allows one to immediately write down how terms in
the Lagrangian are shifted under the change of variables since it
mimics a chiral transformation.

\begin{figure}[!h]
\centering
  \includegraphics[scale=.75]{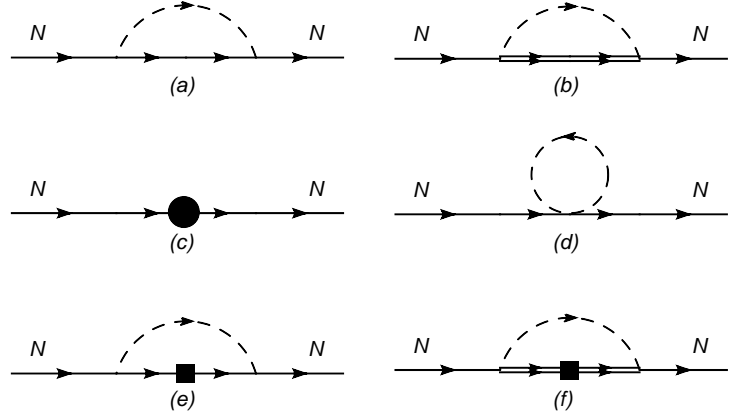}
  \caption{Graphs that contribute to the nucleon mass up to order $\epsilon^4$.
  The square vertex is an insertion of the kinetic energy operator while the
  circle is the direct zero mode mass contribution.}\label{Nmassgraphs}
\end{figure}

In applying the method described above for integrating over the zero
mode portion of the pion fields, care must be taken in the use of
the $V$ and $V^{\dagger}$ matrices. Specifically, one must be sure
that the expansion in powers of $\hat{M}$ has been taken far enough
to include all of the appropriate zero modes in the calculation. A
good example of this is the calculation of the nucleon mass in the
$\epsilon$ regime performed in ref. \cite{Bedaque:2004dt}. To order
$\epsilon^{4}$ the graphs that contribute to the nucleon mass are
given in fig. (\ref{Nmassgraphs}). In fig. \ref{Nmassgraphs}(a),
\ref{Nmassgraphs}(b),
 \ref{Nmassgraphs}(e), and \ref{Nmassgraphs}(f),
the pion couplings are due to one pion derivative interactions
from the $g_{A}^{(0)}$ and $g_{\Delta N}$ Lagrangian terms, so to produce
these graphs one needs only $V=\mathbf{1}$.

The contribution from fig. (\ref{Nmassgraphs}(d)) arises from the
terms in the higher order Lagrangian in eqn. (\ref{1/m lagrangian})
proportional to $\mathcal{A}_{0}^2$ and $\mathcal{A}_{\mu}\mathcal{A}^{\mu}$. Under the change of
variables these become $\mathcal{A}_{\mu}\mathcal{A}^{\mu}\to V \hat{\mathcal{A}}_{\mu}\hat{\mathcal{A}}^{\mu}
V^{\dagger}\sim
V(\partial^{\mu}\hat{\pi}\partial_{\mu}\hat{\pi})V^{\dagger}$, where
again only $V=\mathbf{1}$ is required. Finally at this order in $\epsilon$
is fig. (\ref{Nmassgraphs}(c)), which stems from the Lagrangian term
\begin{eqnarray}
    \lambda C_{1}\bar{N}{ \rm Tr}[m_{q}\Sigma^{\dagger}+h.c.]N
    &\to&\lambda C_{1}\bar{N}{ \rm
    Tr}[m_{q}U^{\dagger}\hat{\Sigma}^{\dagger}U^{\dagger}+h.c.]N\nonumber\\
    &=&\lambda C_{1}\bar{N}{ \rm Tr}[m_{q}(U^{\dagger})^2+h.c.]N+...
\end{eqnarray}
which explicitly carries zero mode information. This term is
evaluated in the same manner as the correction to the pion mass and
results in\cite{Bedaque:2004dt}
\begin{eqnarray}
    m_{B}&\to&
    m_{B}-4m_{\pi}^{2}C_{1}\frac{X'(s)}{2X(s)}\nonumber\\
    X(s)&=&\frac{I_{1}(2s)}{s}
\end{eqnarray}
where $s=\frac{1}{2}m_{\pi}^2f^2V$ and the $I_{n}(s)$'s are modified
Bessel functions. The pion mass also has a shifted value due to
zero-mode contributions such that $m_{\pi}^2 \rightarrow
m_{\pi}^2(s)$\cite{Bedaque:2004dt}, where $m_{\pi}^2(s)=m_{\pi}^2
X'(s)/2X(s)$. Therefore all $m_{\pi}^2$'s appearing in the
computation of the matrix element must be replaced by
$m_{\pi}^2(s)$. All of these results are identical to those in ref.
\cite{Bedaque:2004dt}.

\begin{figure}[!h]
\centering
  \includegraphics[scale=.75]{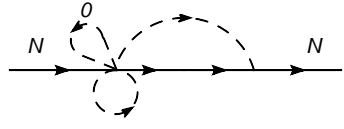}
  \caption{One graph that contributes to the nucleon mass at order $\epsilon^5$.
  An arbitrary number of zero mode loops can exist at the $\bar{N}(\hat{\pi}^3)N$ vertex.}\label{NmassgraphsHO}
\end{figure}

To order $\epsilon^4$ the method presented above coincides with that
of ref. \cite{Bedaque:2004dt} and gives the same result. However, if
one wished to carry out the calculation beyond
$\mathcal{O}(\epsilon^4)$ important differences will surface. An
inspection of the original Lagrangian will lead to graphs with
multiple zero mode loops emanating from the $\pi NN$ vertex and
similar vertices. Fig. (\ref{NmassgraphsHO}) displays an example of
a graph that will contribute at $\mathcal{O}(\epsilon^5)$ to the
nucleon mass (only one zero mode loop is drawn, though an arbitrary
number may exist as they will all be of the same order). In ref.
\cite{Bedaque:2004dt} some of this information does not exist. While
correct to the order they worked, the field redefinition used in
ref. \cite{Bedaque:2004dt} will be incorrect beyond
$\mathcal{O}(\epsilon^4)$. The inconsistency resides in the field
redefinitions $\xi_{L}$ and $\xi_{R}$. The labels of $L$ and $R$ are
not to be taken literally, but rather are a device to recall the two
different ways $\xi$ can transform in order to build chirally
invariant Lagrangian terms. What is true is that there is only one
field, namely $\xi$, such that $\xi^2=\Sigma$. This must hold true
both before and after the field redefinitions. Initially, the
$\Sigma$ field can be thought of as a generic point on the SU(2)
manifold that is a finite distance away from the identity. Defining
$\xi^2=\Sigma$ implies that $\xi$ is located on the midway point of
the geodesic connecting $\Sigma$ to the identity. The field
redefinitions of $\xi_{L} \to U\hat{\xi_{L}}$ and $\xi_{R} \to
U^{\dagger}\hat{\xi_{R}}$ lose this interpretation. The only way to
sensibly interpret the meaning of $\sqrt{U\hat{\Sigma}U}$ is to
appeal to the expansion given in eqn. (\ref{sqrtsigma}).

Turning to the evaluation of the matrix element of the axial-vector
current, the part of the action containing the sources is expanded keeping
only the linear term in each. The matrix element to be evaluated is
then of the form:
\begin{equation}
<N_c(p)|j_{\mu
5}^a|N_{b}(p)>=\frac{1}{\mathcal{Z}_0}\int[\mathcal{D}\Sigma][\mathcal{D}N]
\ N_b^{\dagger}(p) \ j_{\mu 5}^a(q) \ N_c(p) \ e^{-\int d^4x \
\mathcal{L}}
\end{equation}
where $\mathcal{Z}_0$ is the partition function free of sources and
$b$ and $c$ denote particular choices of nucleons from $N^{T}=(p,n)$
in the initial and final states.

\begin{figure}[!h]
\centering
  \includegraphics[scale=.75]{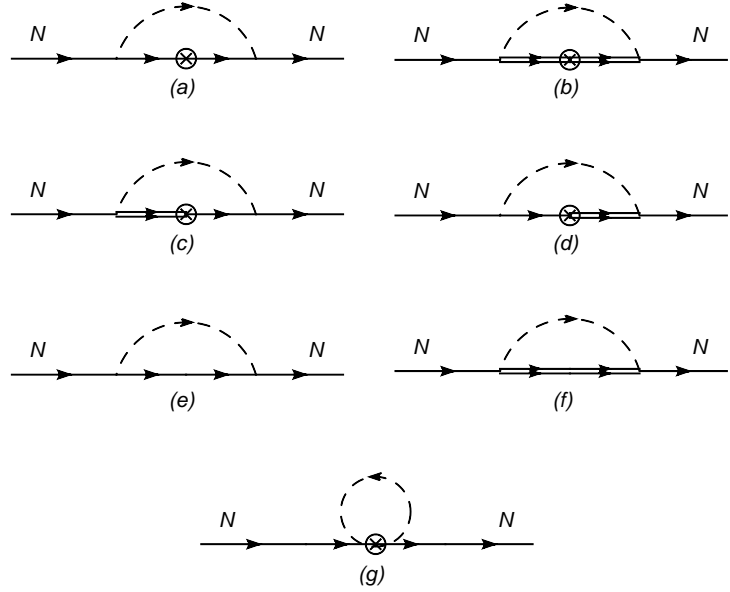}
  \caption{Graphs that contribute at $\mathcal{O}(\epsilon^2)$ and $\mathcal{O}({\epp}^2)$.}\label{epsilongraphs}
\end{figure}

The graphs in fig. (\ref{epsilongraphs}) represent the perturbative
contribution to the matrix element and each of these graphs has a
function of the zero modes that accompany it which must be
integrated over. As an example of how to integrate over the zero
modes, consider the tree level contribution:
\begin{eqnarray}
<N_c(p)|j_{\mu
5}^a|N_{b}(p)>_{tree} \ &=&g_{A}^{(0)}(2\bar{N}S_{\mu}N)(\tau^a)_{mn} \ \frac{\int [\mathcal{D}U^2] \
U_{bm}(U^{\dagger})_{nc} \ {\rm exp}\left(s {\rm Re
Tr}(U^2)\right)}{\int [\mathcal{D}U^2] \ {\rm exp}\left(s {\rm Re
Tr}(U^2)\right)} \label{zeromodeint} \\
&=&g_{A}^{(0)}Q(s)(2\bar{N}S_{\mu}N) \label{zeromodeintanswer}
\end{eqnarray}
where $Q(s)=\frac{1}{3}\left(1+2\frac{I_2(2s)}{I_1(2s)}\right)$.
Here, the $N$'s and $\bar{N}$ are the spinors satisfying
$\frac{1}{2}\left(1+v^{\mu}\gamma_{\mu}\right)N=N$ (with a similar
relationship for $\bar{N}$) not to be confused with the nucleon
field. In the $s \to \infty$ limit one recovers the expected answer
at tree-level, $g_{A}^{(0)}$.

\begin{figure}[!h]
\centering
  \includegraphics[scale=.75]{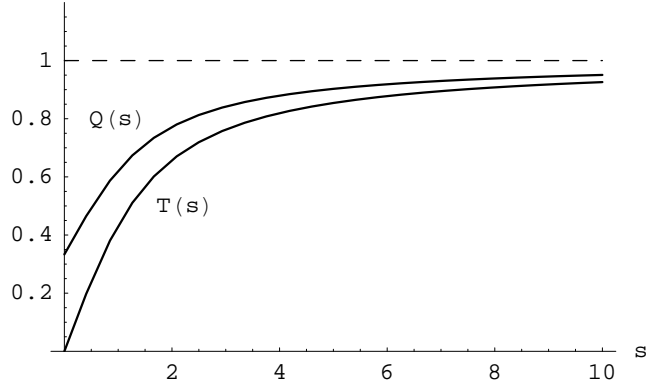}
  \caption{Zero mode factors $Q(s)$ and $T(s)$.}\label{epsiloncoupling}
\end{figure}

A contribution which enters at $\mathcal{O}(\epsilon^2)$ is the
one-loop contribution from the axial-vector current as shown in fig.
(\ref{epsilongraphs}(g)). One can determine the function of $s$ that
will modify the graphs from the zero mode integration. The result
is:
\begin{eqnarray}
    T(s)&=&\frac{I_{0}(2s)}{I_{1}(2s)}-\frac{1}{s}.
\end{eqnarray}
The other contributions at this order are all multiplied by $Q(s)$.
The functions $Q(s)$ and $T(s)$ are plotted in fig.
(\ref{epsiloncoupling}).

\section{Results}
Putting together all of the contributions from the graphs in figure
(\ref{epsilongraphs}) and using the results from appendix A the
result for the $\epsilon$ regime to $\mathcal{O}(\epsilon^3)$ is:
\begin{eqnarray}\label{epsresult}
    \Gamma_{\epsilon}(s)&=&g_{A}^{(0)}Q(s)-\frac{Q(s)}{f^2 L^3}\left[g_{A}^{(0)}\frac{T(s)}{Q(s)}\frac{c_{1}L}{2\pi}
    +\left(\frac{100}{243}g_{\Delta\Delta} \ g_{\Delta N}^{2}+\frac{4}{3}g_{A}^{(0)} \ g_{\Delta N}^{2}\right)
    \left(\frac{c_{1}L}{4\pi}-\frac{c_{2}\Delta L^2}{4\pi^2}+\frac{c_0}{2m_B}\right)\right.\nonumber\\
    &&\left.+\frac{4}{3}(g_{A}^{(0)})^3\left(\frac{c_{1}L}{4\pi}+\frac{c_0}{2m_B}\right)
    -\frac{32}{27}g_{A}^{(0)} g_{\Delta N}^2\left(\frac{c_{1}L}{4\pi}-\frac{c_{2}\Delta L^2}{8\pi^2}
    +\frac{c_0}{2m_B}\right)\right].
\end{eqnarray}
In achieving this result it has been demonstrated in appendix A that
the finite time direction corrections in the $\epsilon$ regime do
not enter until $\mathcal{O}(\epsilon^6)$, thus significantly
reducing the computational burden when working in the $\epsilon$
regime to orders lower than this.

\begin{figure}[!h]
\centering
  \includegraphics[scale=.75]{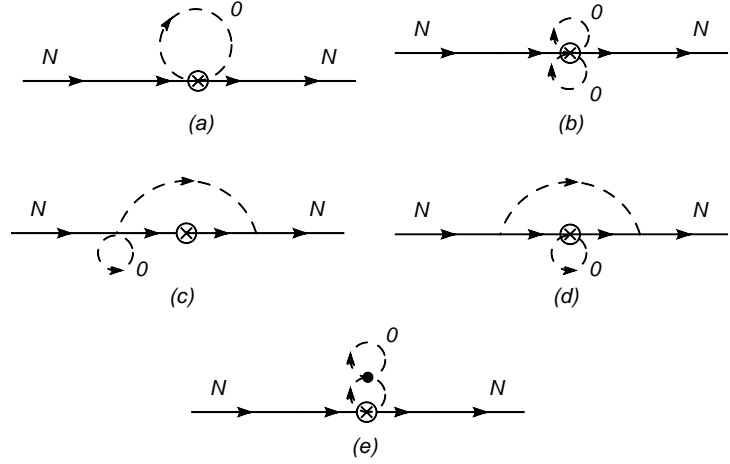}
  \caption{Additional $\epp$ graph contributions. Figure (a) enters at $\mathcal{O}(\epp)$. Figure (b) at $\mathcal{O}(\epp^2)$.
  Figures (c)-(e) are representative of the ways that a zero mode loop can be added to the graphs in fig. \ref{epsilongraphs}, in
  order to create $\mathcal{O}(\epp^3)$ contributions.}\label{eppgraphs}
\end{figure}
The result for the $\epp$ regime utilizes the graphs from fig. (\ref{epsilongraphs})
as well as from fig. (\ref{eppgraphs}). The graphs in fig. (\ref{epsilongraphs})  are
adapted to the $\epp$ regime by letting $s\to\infty$, which gives $Q(s)\to1$ and
$T(s)\to1$, and adding back the diagrams with spatial zero modes:
\begin{eqnarray}\label{eppresult}
    \Gamma_{\epp}&=&\Gamma_{\epsilon}(s=\infty)+\Gamma_{1}+\Gamma_{2}+\Gamma_{3}\nonumber\\
    \Gamma_{1}&=&-\frac{g_{A}^{(0)}}{f^2}\frac{1}{m_{\pi}L^3}\nonumber\\
    \Gamma_{2}&=&\frac{g_{A}^{(0)}}{4f^4}\frac{1}{m_{\pi}^{2}L^6}\nonumber\\
    \Gamma_{3}&=&\frac{1}{2f^4}\frac{1}{m_{\pi}L^6}\left[-\frac{g_{A}^{(0)}}{20}\frac{1}{m_{\pi}^{2}L^3}
    +g_{A}^{(0)}\frac{c_{1}L}{4\pi}-\frac{320}{81}g_{A}^{(0)}g_{\Delta N}^2 \left(\frac{c_{1}L}{4\pi}-\frac{c_{2}\Delta L^2}{8\pi^2}
    +\frac{c_0}{2m_B}\right)\right.\nonumber\\
    &&\left.+\frac{40}{9}(g_{A}^{(0)})^{3}\left(\frac{c_{1}L}{4\pi}+\frac{c_0}{2m_B}\right)
    +\left(\frac{1000}{729}g_{\Delta N}^{2} \ g_{\Delta\Delta}
    +\frac{40}{9}g_{\Delta N}^2 \ g_{A}^{(0)}\right)\left(\frac{c_{1}L}{4\pi}
    -\frac{c_{2}\Delta L^2}{4\pi^2}+\frac{c_0}{2m_B}\right)\right]\nonumber\\
    &&+\frac{8}{3}\frac{1}{f^4}\frac{1}{m_{\pi}L^6}\left(g_{A}^{(0)}\frac{c_{1}L}{4\pi}
    +(g_{A}^{(0)})^{3}\left(\frac{c_{1}L}{4\pi}+\frac{c_0}{2m_B}\right)\right.\nonumber\\
    &&\left.+\left(\frac{10}{27}g_{\Delta\Delta} \ g_{\Delta N}^{2}
    +\frac{10}{9}g_{A}^{(0)} \ g_{\Delta N}^2 \right)
    \left(\frac{c_{1}L}{4\pi}-\frac{c_{2}\Delta L^2}{4\pi^2}+\frac{c_0}{2m_B}\right)\right.\nonumber\\
    &&\left.-\frac{8}{9}g_{A}^{(0)} \ g_{\Delta N}^2 \left(\frac{c_{1}L}{4\pi}
    -\frac{c_{2}\Delta L^2}{8\pi^2}+\frac{c_0}{2m_B}\right)\right)
\end{eqnarray}
where the subscripts refer to the order in $\epp$ considered. In
both the $\epsilon$ and $\epp$ results given above expansions in
$m_{\pi}L/2\pi$ and $\Delta L/2\pi$ have been taken as these
quantities count as order $\epsilon$ and $\epp$ respectively.

\section{Discussion and Conclusions}
In this paper we have discussed the various calculational regimes
that are relevant to current lattice calculations. In fig.
(\ref{etaregimes}) the approximate extent of these regimes have been
demonstrated for different volumes and pion masses, as calculated
using the tadpole diagram as a guide. However, it was stressed that
these outlines are not to be interpreted as strict boundaries, but
rather as midpoints in the smooth transition between the different
regimes. From the figure it is clear that as one moves toward a more
hypercubical lattice (with the temporal direction the same length as
the spatial extent of the volume) at smaller volumes and more
physical pion masses that the $\epsilon$ and $\epp$ regimes will be
the most important.

Ultimately one would like to use the results from the $\epsilon$ and
$\epp$ regime in eqns. (\ref{epsresult}) and (\ref{eppresult}) to
fit to lattice data for $g_{A}$ and extrapolate to the physical pion
mass value. The majority of the data that exists on lattice
calculations of $g_{A}$ can be found in refs.
\cite{Capitani:1999zd,Dolgov:2002zm,
Ohta:2004mg,Khan:2004vw,Edwards:2005ym,Khan:2006de}. Most of these
calculations of the axial current have been performed in the $\epp$,
$\delta$, or $p$ regimes according to the separation given above.
While there is a significant amount of data available from these
calculations, in order to avoid working in the $\delta$ regime as
defined above, only results from volumes significantly greater than
$(2 \ {\rm fm})^3$ should be used. However, the number of points
that strictly satisfy this condition, as well as the condition from
eqn. (\ref{pregimebound}) that $m_{\pi}L/2\pi<1$, is too few for any
reasonable fit to be performed.

In ref. \cite{Edwards:2005ym} a fit to the LHPC Collaboration
lattice data was performed using a $p$ regime calculation of
$g_{A}$, as most of these points fall within that regime. The most
effective way to efficiently reduce the statistical error in the
extrapolation of $g_{A}$ at the physical point (where one wishes to
compare to experiment) is to perform lattice calculations at lower
pion mass ($<350$ MeV). Generally speaking, one low mass calculation
may easily be equivalent to three or four high mass calculations in
terms of ability to reduce the statistical error of the
extrapolation. While performing lower pion mass lattice calculations
is often computationally expensive, increases in the speed and
number of computers available will soon put these low mass
calculations within reach. In addition, a high pion mass moves one
further away from the regime of ChPT, adding to statistical
considerations concerns about the applicability of the effective
field theory, regardless of the regime one is working in. Clearly,
the way forward involves lattice calculations at lower pion mass,
where the $\epsilon$ and $\epp$ regimes will become increasingly
important. Our results allow coherent analysis of such data.

In the $\epsilon$ regime the technique of collective variables was
used to separate the spacetime independent zero modes from the
spacetime dependent non-zero modes. Before the split into zero and
non-zero modes, one is able to see that in graphs where non-zero
modes are connected to external states one can have arbitrary
numbers of zero mode loops present. These contributions are not lost
in the field redefinition. To usefully separate out the order one
components of a given contribution it was demonstrated that one must
perform the change of variables that is defined by
$\Sigma=U\hat{\Sigma}U$ as in eqn. (\ref{chgofvar}). To correctly
include baryon interactions after this change of variables the $\xi$
objects must be treated carefully with the change of variables given
in eqns. (\ref{xichgvars1}) and (\ref{xichgvars2}). We have
described a method for doing this that differs from previous
methods\cite{Bedaque:2004dt}.

\begin{figure}[!h]
\centering
  \includegraphics[scale=.75]{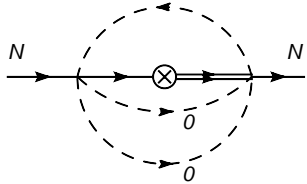}
  \caption{A diagram for which an integration over the zero modes does not
  correspond to a simple redefinition of coupling constants.}\label{zerodiagram}
\end{figure}

One feature of the method proposed for dealing with the
$\mathcal{O}(1)$ contributions of the pion fields is that the
integration over the zero modes must be performed on a diagram by
diagram basis. It may seem legitimate to integrate out the zero mode
contributions and return a Lagrangian where the relevant coupling
constants have been replaced by the appropriate functions of $s$,
that have been calculated "a posteriori," and with all insertions of
the pion fields containing purely non-zero mode information. Doing
this, however, would be incorrect. Evaluating higher order diagrams
will lead to different zero mode functions multiplying their
respective non-zero mode Feynman diagrams. An example of such a
diagram is given in fig. (\ref{zerodiagram}). These zero mode
functions, in general, will not be encoded in a coupling constant
redefinition; it would be misleading to label $g_{A}$ as $g_{A}(s)$.
As a result, one cannot write down a local Lagrangian after the zero
modes have been integrated out. To perform the actual integrations
on low order diagrams with relatively simple zero mode contributions
one can either use hyperspherical coordinates or use the symmetry
properties of the SU(2) group to obtain an answer. However, with
more difficult and complicated higher order diagrams hyperspherical
coordinates are the most effective tool, as shown in appendix C.

To conclude, the matrix element of the nucleon axial-vector current
has been calculated within the $\epsilon$ and $\epp$ regimes to
$\mathcal{O}(\epsilon^3)$ and $\mathcal{O}({\epp}^3)$ respectively.
The $\epp$ regime result agrees to NLO with previous
calculations\cite{Detmold:2004ap}. A method was put forward to
calculate the nonperturbative contributions of pion zero momentum
modes in the $\epsilon$ regime in the presence of baryons. This
method can be used in any calculation, diagram by diagram, at all
orders in the $\epsilon$ expansion. In calculating $g_{A}$ it was
discovered that the temporal direction could be approximated as
infinite, with the errors introduced entering only at the
$\mathcal{O}(\epsilon^6)$ level, as shown in appendix A. This
significantly reduces the computational burden of calculations in
the $\epsilon$ regime.

\acknowledgments We are very grateful to W. Detmold and M. Savage
for their time and many useful conversations. In addition, B.S. and
J.W. are indebted to A. Iqbal, D. B. Kaplan, S. Sharpe, M. Endres,
C. Kozcaz, C. Spitzer, and A. Walker-Loud for comments and
discussions.

\appendix
\section*{Appendix}
\renewcommand{\theequation}{A-\arabic{equation}}
\setcounter{equation}{0}
\section{Integral \& Sum Manipulations}
The integrals used in the calculation of the results in Euclidean space are:
\begin{equation} \label{1loopint}
    R(m_\pi)=\mu^{d-4}\int\frac{d^{d}q}{(2\pi)^d}\frac{1}{q_{0}^{2}+\vec{q}^{2}+m_{\pi}^{2}}
\end{equation}
\begin{equation}\label{2bardecint}
    J(\Delta ,
    m_\pi,\beta,L,m_B)=\mu^{d-4}\int\frac{d^{d}q}{(2\pi)^d}\frac{|\vec{q}|^2}{(iq_{0}-\Delta+\vec{q}^{2}/2m_B)^2}\frac{1}{q^2+m_{\pi}^2}
\end{equation}
\begin{eqnarray}\label{LRbardecint}
    N(\Delta,m_\pi,\beta,L,m_B)&=&\mu^{d-4}\int\frac{d^{d}q}{(2\pi)^d}\frac{1}{q^2+m_{\pi}^{2}}
    \frac{1}{iq_{0}-\Delta+\vec{q}^{2}/(2m_B)}\frac{|\vec{q}|^2}{iq_{0}+\vec{q}^{2}/(2m_B)}
\end{eqnarray}
In the $\epp$ regime one normally integrates over the $q_0$
component and forms a sum over the remaining finite volume spatial
components, while for the $\epsilon$ regime a sum over all four
components would be formed. However, the $\epp$ regime can be
recovered as a limit of the $\epsilon$ finite volume manipulations
and so one can focus on those.

The sums that arise during our calculation are defined within a
dimensional regularization framework so that these should merge smoothly with
the infinite volume pieces as $L\to\infty$. Toward this end the
sums must be defined by the relation
\begin{equation}
    \frac{1}{L^3}\sum_{\vec{q}}=\frac{1}{L^3}\sum_{\vec{q}}^{\Lambda}-\int^{\Lambda}\frac{d^{3}q}{(2\pi)^3}
    +\mu^{d-4}\int^{\rm{DR}}\frac{d^{d-1}q}{(2\pi)^{d-1}}
\end{equation}
where the cutoff dependence of the sum is removed by that of the
integral. With this definition, the dimensionally regularized
integrals will be the infinite volume pieces of the diagram while
the remaining sums are the finite volume corrections. The DR pieces
will all be proportional to $m_{\pi}^2$ or $\Delta^2$ and so will
technically enter at order $\epsilon^4$ or ${\epp}^4$. However, in
order to smoothly merge with the infinite volume result these higher
order terms must be included along with a $\mu$ dependent
counterterm. These DR pieces are calculated in the normal
way\cite{Beane:2004rf}.

For the temporal sums found in the $\epsilon$ regime, the Abel-Plana formula
is needed and is given as:
\begin{eqnarray}\label{betasum}
    \frac{1}{\beta}\sum_{n}f\left(\frac{2\pi n}{\beta}\right)&=&\int^{\infty}_{-\infty}\frac{dz}{2\pi}f(z)
    -i{\rm Res}\left(\frac{f(z)}{e^{i\beta z}-1}\right)\mid_{\rm{lower
    plane}}+i{\rm Res}\left(\frac{f(z)}{e^{-i\beta z}-1}\right)\mid_{{\rm upper plane}}
\end{eqnarray}
which is valid if $f(z)$ has no poles on the real axis. The first
term on the right hand side is the infinite time direction piece
used in $\epp$ calculations. One can thus look at eqn.
(\ref{betasum}) as the infinite time $\epp$ piece plus
finite time $\epsilon$ corrections.

Using eqn. (\ref{betasum}) the integral from eqn. (\ref{1loopint}) becomes:
\begin{eqnarray}
    R(m_\pi)&=&\int\frac{d^{4}q}{(2\pi)^4}\frac{1}{q_{0}^{2}+\vec{q}^{2}+m_{\pi}^{2}} \nonumber \\
    &=&\frac{1}{\beta L^3}\sum_{n_\mu}\frac{1}{(2\pi n_0/\beta)^2+(2\pi \vec{n}/L)^2+m_{\pi}^{2}} \nonumber \\
    &=&\frac{1}{L^3}\sum_{\vec{n}}\left(\frac{1}{\beta}\sum_{n_0}\frac{1}{(2\pi n_0/\beta)^2
    +(2\pi \vec{n}/L)^2+m_{\pi}^{2}}\right)
\end{eqnarray}
Now let $\beta=aL$ for some value of $a$ where one expects
$1\le a\le\infty$. It is now possible to expand the above expression in powers
of $m_{\pi}L$ as $m_{\pi}L\sim\epsilon^\alpha$ with $\alpha\ge1$.
Splitting off the zero mode part of this expression and expanding in $m_{\pi}L$ gives
\begin{eqnarray}\nonumber
    R_{0}(m_\pi,\beta,L)&=&\frac{1}{2}\frac{1}{m_{\pi}L^3}\\
    R_{\emptyset}(m_\pi,\beta,L)&=&\frac{1}{L^3}\sum_{\vec{n}\not=\vec{0}}\frac{L}{4\pi}\frac{1}{|\vec{n}|}
    +\frac{1}{L^3}\sum_{\vec{n}\not=\vec{0}}\left(\frac{L}{2\pi}\frac{1}{|\vec{n}|}
    \frac{1}{e^{2\pi a|\vec{n}|}-1}\right)+\mathcal{O}(\epsilon^4)\nonumber
\end{eqnarray}
where the infinite time direction term is the first term on the
right and the finite time correction is the second term.
Looking at the sums present in each term, the finite time
correction will converge very quickly due to the presence
of the exponential. In fact, even for just the first term in the sum
the finite time piece will be suppressed by a factor of
267 under the most favorable conditions ($a=1$). Given that
current lattice calculations return
$\epsilon>1/4$, the sum for the finite time corrections of
the one pion loop is effectively at $\epsilon^4$ or higher, and so
for the $\epsilon^2$ single pion loop, the finite time
corrections are effectively at $\epsilon^6$ and can be safely
ignored. Thus:
\begin{eqnarray}
    R_{0}(m_\pi,\beta,L)&=&\frac{1}{2}\frac{1}{m_{\pi}L^3}\\
    R_{\emptyset}(m_\pi,\beta,L)&=&\frac{1}{L^3}\sum_{\vec{n}\not=\vec{0}}\frac{L}{4\pi}\frac{1}{|\vec{n}|}
    +\mathcal{O}(\epsilon^4)
\end{eqnarray}
for both the $\epsilon$ and $\epp$ regimes.

For eqns. (\ref{2bardecint}) and (\ref{LRbardecint}) the same
procedure is used that was used on eqn. (\ref{1loopint}) with the
only modifications being that: (1) the spatial zero modes vanish due
to the integrand's proportionality to $|\vec{q}|^2$, and (2) the
presence of the baryon mass in eqn. (\ref{baryonfourier}) must be
used in computing the finite volume residues. The pole is in the
lower half plane and will give a finite time correction of
\begin{equation}
    -i{\rm Res}\left(\frac{f(z)}{e^{i\beta
    z}-1}\right)|_{{\rm lower}}\propto\frac{1}{e^{\beta(\Delta+m_B)}-1}\approx0
\end{equation}
as $m_{B}\sim\Lambda_{\chi} $\cite{Bedaque:2004dt}.
The nucleon sector will give the
same result with $\Delta\to0$. Because of this fact which arose due
to the finite volume boundary conditions on the relativistic field from eqn. (\ref{baryonfourier}),
the finite time corrections due to the nucleon and decuplet
poles do not contribute. From the above results, the finite time
corrections from the pion propagator poles do not contribute
at the order we work. Therefore, to order $\mathcal{O}(\epsilon^3)$, the finite
time contributions to the sums are negligible and only the infinite
time direction contributions are needed.

Calculating eqn. (\ref{2bardecint}) explicitly and expanding in $m_{\pi}L$ and $\Delta L$:
\begin{eqnarray}
     J(\Delta ,
    m_\pi,\beta,L,m_B)
    &=&\frac{1}{\beta L^3}\sum_{n_\mu}
    \frac{(2\pi\vec{n}/L)^2}{(i(2\pi n_0/\beta)-\Delta+(2\pi\vec{n}/L)^2/2m_B)^2}\frac{1}{(2\pi n_0/\beta)^2+(2\pi\vec{n}/L)^2+m_{\pi}^2}\nonumber\\
    &=&\frac{1}{L^3}\sum_{\vec{n}\not=0}\left(\frac{L}{4\pi}\frac{1}{|\vec{n}|}
    -\frac{\Delta L^2}{4\pi^2}\frac{1}{|\vec{n}|^2}+\frac{1}{2m_B}\right)+\mathcal{O}(\epsilon^4)
\end{eqnarray}
For eqn. (\ref{LRbardecint}) it is easier to generalize the integral:
\begin{eqnarray}
    N'(A,B,m_\pi,\beta,L,m_B)&=&\int\frac{d^{4}q}{(2\pi)^4}\frac{1}{q^2+m_{\pi}^{2}}
    \frac{1}{iq_{0}-A+q^{2}/(2m_B)}\frac{|\vec{q}|^2}{iq_{0}-B+q^{2}/(2m_B)}\\
    N(\Delta,m_\pi,\beta,L,m_B)&=&N'(\Delta,0,m_\pi,\beta,L,m_B)
\end{eqnarray}
Using similar methods to those used to evaluate the function $J$ and
taking $A\to\Delta$ and $B\to0$ gives
\begin{eqnarray}
    N(\Delta,m_\pi,\beta,L,m_B)&=&\frac{1}{L^3}\sum_{\vec{n}\not=0}\left(\frac{L}{4\pi}\frac{1}{|\vec{n}|}
    -\frac{\Delta L^2}{8\pi^2}\frac{1}{|\vec{n}|^2}+\frac{1}{2m_B}\right)+\mathcal{O}(\epsilon^4)
\end{eqnarray}

For each of the resulting equations one is left with a spatial sum
over various powers of $|\vec{n}|$. Several of these sums contain
divergent pieces which must be removed and placed in appropriate
counterterms. Defining these sums as the limit of an analytically
continued sum from non-integer powers of $|\vec{n}|$, one can indeed
obtain the finite pieces that are necessary. This results
in\cite{Edery:2005bx, Luscher:1986pf}:
\begin{eqnarray}
    c_2&=&\sum_{\vec{n}\not=0}\frac{1}{|\vec{n}|^2}=-8.913633\nonumber\\
    c_1&=&\sum_{\vec{n}\not=0}\frac{1}{|\vec{n}|}=-2.8372974\nonumber\\
    c_0&=&\sum_{\vec{n}\not=0}1=-1\nonumber\\
\end{eqnarray}
With these coefficients the sums above become
\begin{eqnarray}
    R_{0}(m_\pi,\beta,L)&=&\frac{1}{2}\frac{1}{m_{\pi}L^3}\\
    R_{\emptyset}(m_\pi,\beta,L)&=&\frac{1}{L^3}\frac{c_{1}L}{4\pi}+\mathcal{O}(\epsilon^4)\\
    J(\Delta ,m_\pi,\beta,L,m_B)&=&\frac{1}{L^3}\left(\frac{c_{1}L}{4\pi}
    -\frac{c_{2}\Delta L^2}{4\pi^2}+\frac{c_{0}}{2m_B}\right)+\mathcal{O}(\epsilon^4)\\
    N(\Delta,m_\pi,\beta,L,m_B)&=&\frac{1}{L^3}\left(\frac{c_{1}L}{4\pi}
    -\frac{c_{2}\Delta L^2}{8\pi^2}+\frac{c_{0}}{2m_B}\right)+\mathcal{O}(\epsilon^4)
\end{eqnarray}

\renewcommand{\theequation}{B-\arabic{equation}}
\setcounter{equation}{0}
\section{Contributing Diagrams}
\subsection{Order $\epp$ Graphs}
For the graph in fig. \ref{eppgraphs}(a) the
zero next to the pion loop indicates that we are only interested in
the zero momentum mode of this pion loop. This graph contributes to
$\mathcal{O}(\epp)$. Its contribution is
\begin{eqnarray}
    \Gamma_{1}&=&-\frac{2g_{A}^{(0)}}{f^2}R_{0}(m_\pi,\beta,L)\bar{N}S^{\mu}N
    \label{oneloopcurrent}
\end{eqnarray}

\subsection{Order $\epsilon^2$ and ${\epp}^{2}$ Graphs}
The graphs in fig. \ref{epsilongraphs}(a)-(d)
yield\cite{Detmold:2004ap}:
\begin{eqnarray}\label{e2twobar}
    \Gamma_{2a}&=&\frac{1}{6f^2} \ (g_{A}^{(0)})^{3} \ Q(s) \ J(0,m_\pi,\beta,L,m_B) \ \bar{N}S^{\mu}N \\
    \Gamma_{2b}&=&-\frac{100}{243}\frac{g_{\Delta\Delta}g_{\Delta N}^2 Q(s)}{f^2}J(\Delta,m_\pi,\beta,L,m_B)\bar{N}S^{\mu}N \\
    \Gamma_{2c}&=&\Gamma_{2d}=\frac{32}{27}\frac{g_{A}^{(0)}g_{\Delta N}^2 Q(s)}{f^2}N(\Delta,m_\pi,\beta,L,m_B)\bar{N}S^{\mu}N
\end{eqnarray}

The two graphs which contribute to field renormalization are given in figs.
\ref{epsilongraphs}(e) and (f).
\begin{eqnarray}
    \Gamma_{2e}&=&-\frac{3}{2}\frac{(g_{A}^{(0)})^3 Q(s)}{f^2}J(0,m_\pi,\beta,L,m_B)\bar{N}S^{\mu}N \\
    \Gamma_{2f}&=&-\frac{4}{3}\frac{g_{A}^{(0)}g_{\Delta N}^2 Q(s)}{f^2}J(\Delta,m_\pi,\beta,L,m_B)\bar{N}S^{\mu}N
\end{eqnarray}
Figure \ref{epsilongraphs}(g) gives:
\begin{equation}
    -\frac{2g_{A}^{(0)}T(s)}{f^2}R_{\emptyset}(m_\pi,\beta,L)\bar{N}S^{\mu}N \\
\end{equation}
and fig. \ref{eppgraphs}(b) give a contribution of:
\begin{equation}
     \frac{g_{A}^{(0)}}{f^4}[R_{0}(m_\pi,\beta,L)]^{2}\bar{N}S^{\mu}N
\end{equation}
where the function $R_{\emptyset}(m_\pi,\beta,L)$ is the non-zero
mode portion of the function $R(m_\pi)$.

\subsection{Order $\epsilon^3$ and ${\epp}^3$ Graphs}
At order ${\epp}^3$, there exist contributions from graphs such as
those shown in fig. \ref{eppgraphs}(c) and (d). To generate these graphs one
adds a spatial zero mode loop to each vertex in fig. \ref{epsilongraphs}(a). As
these contain spatial zero mode pion loops they contribute only to
$\mathcal{O}({\epp}^3)$ and not to $\mathcal{O}({\epsilon}^3)$.
Adding a zero mode pion loop to the $\mathcal{O}(\epsilon^2)$
graphs at every vertex on every graph, including the field
renormalization graphs, will generate the majority of the $\mathcal{O}(\epp^3)$ contributions.
The effect of adding these pion zero mode loops will be to change the coefficient due to that vertex and
multiply the previous expressions by the pion loop integral
$R_0(m_\pi)$. None of the Clebsch-Gordan coefficients or the spin
operator contractions will change from those used at
$\mathcal{O}({\epp}^2)$. To find the change to the vertex
coefficients one needs to look at each vertex in turn.

By carefully expanding the vertex terms in the Lagrangian to
three pions one can derive the change that occurs to the
graphs found when a zero mode pion
loop is added to  figs. (\ref{epsilongraphs}(a)-(f)). Upon
completing the expansion to three pions, such a
change simply multiplies the graphs from $\mathcal{O}({\epp}^2)$ by a
factor of $-\frac{2}{3f^2}R_{0}(m_\pi,\beta,L)$. In addition, one can add
such a zero mode loop to either end of the graph so
that there are two distinct configurations. Hence, adding a zero
mode pion loop to the Lagrangian vertex will give an $\mathcal{O}(\epp^3)$
contribution that looks like
\begin{eqnarray}\label{vertexzero}
    \Gamma_{3a}&=&\left(-\frac{4}{3f^2}R_{0}(m_\pi,\beta,L)\right)\left(-\frac{1}{f^2}
    \left[\frac{100}{243}g_{\Delta N}^{2}g_{\Delta\Delta}J(\Delta,m_\pi,\beta,L,m_B)\right.\right. \nonumber \\
    &&\left.\left.+\frac{4}{3}g_{A}^{(0)}g_{\Delta N}^{2}J(\Delta,m_\pi,\beta,L,m_B)+\frac{4}{3}(g_{A}^{(0)})^3 J(0,m_\pi,\beta,L,m_B)\right.\right.\nonumber\\
    &&\left.\left.-\frac{32}{27}g_{A}^{(0)}g_{\Delta N}^2 N(\Delta,m_\pi,\beta,L,m_B)\right]\right)\bar{N}S^{\mu}N
\end{eqnarray}

$\mathcal{O}(\epp^3)$ diagrams are also produced by adding a
zero mode pion loop to the vertex due to the current for each of the
graphs in fig. (\ref{epsilongraphs}) as well as the contribution from fig. \ref{eppgraphs}(b).
This includes adding a zero mode loop to the current vertex that
is implicitly part of the field renormalization graphs, creating an $\mathcal{O}(\epp^3)$
contribution that effectively combines the contribution from fig. $\ref{eppgraphs}(a)$
with the field renormalizations of figs. \ref{epsilongraphs}(e) and (f). Figures \ref{epsilongraphs}(a)-(f)
are multiplied by a factor
of $-\frac{2}{f^2}R_{0}(m_\pi,\beta,L)$, while fig.
\ref{epsilongraphs}(g) is multiplied by
$-\frac{1}{2f^2}R_{0}(m_\pi,\beta,L)$, fig.
\ref{eppgraphs}(b) is multiplied by
$-\frac{1}{5f^2}R_{0}(m_\pi,\beta,L)$. Putting all of these current
insertion loops together yields a contribution of
\begin{eqnarray}\label{currentzero}
    \Gamma_{3b}&=&\left(-\frac{1}{5}\frac{g_{A}^{(0)}}{f^6}[R_{0}(m_\pi,\beta,L)]^{3}+\frac{g_{A}^{(0)}}{f^4}R_{0}(m_\pi,\beta,L)R_{\emptyset}(m_\pi,\beta,L)\right.\nonumber\\
    &&\left.-\frac{2}{f^2}R_{0}(m_\pi,\beta,L)\left(-\frac{1}{f^2}
    \left[\frac{100}{243}g_{\Delta N}^{2}g_{\Delta\Delta}J(\Delta,m_\pi,\beta,L,m_B)\right.\right.\right. \nonumber \\
    &&\left.\left.\left.+\frac{4}{3}g_{A}^{(0)}g_{\Delta N}^{2}J(\Delta,m_\pi,\beta,L,m_B)+\frac{4}{3}(g_{A}^{(0)})^3 J(0,m_\pi,\beta,L,m_B)\right.\right.\right.\nonumber\\
    &&\left.\left.\left.-\frac{32}{27}g_{A}^{(0)}g_{\Delta N}^2 N(\Delta,m_\pi,\beta,L,m_B)\right]\right)\right)\bar{N}S^{\mu}N
\end{eqnarray}

The final $\mathcal{O}({\epp}^3)$ graphs are those as in fig.
\ref{eppgraphs}(e), where the vertex denoted by the dot is from an
expansion of the pion kinetic energy term. This will generate a four
pion vertex, and by choosing the correct contractions of the pion
fields a zero mode loop can be added as shown to each of the graphs
in fig. (\ref{epsilongraphs}). This gives a
contribution of
\begin{eqnarray}
    \Gamma_{3c}&=&\frac{32}{3}\frac{g_{A}^{(0)}}{f^4}R_{\emptyset}(m_\pi,\beta,L)\frac{d}{dm_{\pi}^2}R_{0}(m_\pi,\beta,L)
    +\frac{16}{3f^4}R_{0}(m_\pi,\beta,L)\left(2g_{A}^{(0)}\left[R_{\emptyset}(m_\pi,\beta,L)
    +m_{\pi}^{2}\frac{d}{dm_{\pi}^2}R_{\emptyset}(m_\pi,\beta,L)\right]\right.\nonumber\\
    &&\left.+(g_{A}^{(0)})^3 \left[J(0,m_\pi,\beta,L,m_B)+m_{\pi}^{2}\frac{d}{dm_{\pi}^2}J(0,m_\pi,\beta,L,m_B)\right]\right.\nonumber\\
    &&\left.+\left(\frac{10}{27}g_{\Delta\Delta}g_{\Delta N}^{2}+\frac{10}{9}g_{A}^{(0)}g_{\Delta N}^2 \right)\left[J(\Delta,m_\pi,\beta,L,m_B)
    +m_{\pi}^{2}\frac{d}{dm_{\pi}^2}J(\Delta,m_\pi,\beta,L,m_B)\right]\right.\nonumber\\
    &&\left.-\frac{8}{9}g_{A}^{(0)}g_{\Delta N}^2 \left[N(\Delta,m_\pi,\beta,L,m_B)
    +m_{\pi}^{2}\frac{d}{dm_{\pi}^2}N(\Delta,m_\pi,\beta,L,m_B)\right]\right)\bar{N}S^{\mu}N
\end{eqnarray}

\begin{figure}[!h]
\centering
  \includegraphics[scale=.75]{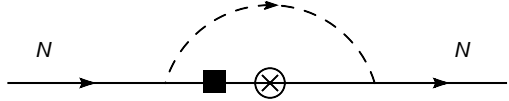}
  \caption{Example of operator insertions from the $\mathcal{O}(1/m_B)$ Lagrangian. This graph would contribute at
  $\mathcal{O}({\epp}^3)$ and $\mathcal{O}({\epsilon}^3)$ but is already counted by the choice of propagator.}\label{e3mBgraph}
\end{figure}

From an operator point of view, both the $\epsilon$ and $\epp$
regimes have graphs of $\mathcal{O}(\epsilon^3)$ and
$\mathcal{O}({\epp}^3)$ that look like that in fig.
(\ref{e3mBgraph}). The square vertex is an insertion of the operator
from the $\mathcal{O}(1/m_B)$ Lagrangian terms found in eqn.
(\ref{1/m lagrangian}). However, these graphs have been
automatically accounted for by the choice of propagator when the
$\mathcal{O}(1/m_B)$ terms were included.

\renewcommand{\theequation}{C-\arabic{equation}}
\setcounter{equation}{0}
\section{Haar Measure}
The integration measure to be used for the zero mode is
$[\mathcal{D}U^2]$ (parameterized with hyperspherical coordinates):
\begin{eqnarray}
\int [\mathcal{D}U^2]&=&\frac{1}{\pi^2}\int d^4a \ \delta(a^2-1)
\end{eqnarray}
such that $U^2=a_{0} \mathbf{1}+i\vec{a}\cdot\vec{\sigma}$ and the
integral is normalized to one. For the calculations in this paper
$U=b_{0} \mathbf{1}+i\vec{b}\cdot\vec{\sigma}$ is parameterized in
terms of hyperspherical coordinates: $|b| \in [0,\infty]$, $\psi \in
[0,\pi]$, $\theta \in [0,\pi]$, and $\phi \in [0,2\pi]$ giving
\begin{eqnarray}
U&=&|b|{\rm cos}(\psi)\mathbf{1}+i|b|{\rm sin}(\psi){\rm
sin}(\theta){\rm cos}(\phi)\sigma_1+i|b|{\rm sin}(\psi){\rm
sin}(\theta){\rm sin}(\phi)\sigma_2+i|b|{\rm sin}(\psi){\rm
cos}(\theta)\sigma_3
\end{eqnarray}
In order to express the integration measure in terms of the radial
and angular coordinates describing $U$ we merely have to calculate a
Jacobian factor. The result obtained is:
\begin{eqnarray}
\int [\mathcal{D}U^2] &\to& \frac{1}{4\pi^2}\int d^4b \ (16b^4 {\rm cos}^2(\psi)) \ \delta(b^2-1) \nonumber \\
&=&\frac{4}{\pi^2}\int db \ b^7 \ \delta(b^2-1) \int d\psi \ d\theta
\ d\phi \ {\rm sin}^2(\psi) {\rm cos}^2(\psi){\rm sin}(\theta)
\end{eqnarray}
where in the first line the prefactor of $1/4$ is present so the
integral is properly normalized and the parenthetical factor is the
Jacobian.

The integral in the numerator of eqn. (\ref{zeromodeint}) can be
done by parameterizing the zero mode variable, $U$, in terms of
hyperspherical coordinates or in this particular example it can be
done directly. To see this, the integrand and measure in eqn.
(\ref{zeromodeint}) must be invariant under multiplication on the
left and right by arbitrary constant SU(2) matrices. This implies
that the integral must have the form:
\begin{eqnarray}
\int [\mathcal{D}U^2] \ U_{bm}(U^{\dagger})_{nc} \ {\rm exp}\left(s {\rm Re Tr}(U^2)\right)&=&A(s)\delta_{bc}\delta_{mn}+B(s)\delta_{bm}\delta_{nc}
\end{eqnarray}
Contracting indices and using the relation $\int [\mathcal{D}U^2] \
{\rm exp}\left(s {\rm Re Tr}(U^2)\right)=X(s)$\cite{Gasser:1987ah,
Creutz:1984mg}, and $2 \ {\rm det} \ U=({\rm Tr} \ U)^2-{\rm Tr} \
U^2$ (valid for SU(2) matrices) one will obtain:
\begin{eqnarray}
A(s)&=&\frac{1}{6}\left(2X(s)-X^{\prime}(s)\right) \\
B(s)&=&\frac{1}{3}\left(X(s)+X^{\prime}(s)\right)
\end{eqnarray}

\renewcommand{\theequation}{D-\arabic{equation}}
\setcounter{equation}{0}
\section{Matrix Definitions}
The form of the matrix $A$ defined earlier in the paper has the
following structure (using the hyperspherical parametrization from
eqn. (\ref{hypersphere})):
\begin{eqnarray}
A_{11}&=&\frac{1}{\sqrt{2}}\hat{\pi}_0 {\rm cos}(\psi)+\frac{i}{\sqrt{2}}\hat{\pi}_0{\rm cos}(\theta){\rm sin}(\psi)
+\frac{i}{2}e^{-i\phi}\hat{\pi}_{-}{\rm sin}(\theta){\rm sin}(\psi)+
\frac{i}{2}e^{-i\phi}\hat{\pi}_{+}{\rm sin}(\theta){\rm sin}(\psi)\nonumber \\
&&- \frac{i}{2}e^{-i\phi}\hat{\pi}_{-}{\rm cos}(\theta){\rm sin}(\theta){\rm sin}(\psi){\rm tan}(\psi)-
\frac{i}{2}e^{-i\phi}\hat{\pi}_{+}{\rm cos}(\theta){\rm sin}(\theta){\rm sin}(\psi){\rm tan}(\psi)\nonumber \\
&&+\frac{1}{\sqrt{2}}\hat{\pi}_0 {\rm sin}^2(\theta){\rm
sin}(\psi){\rm tan}(\psi)\\
A_{12}&=&\hat{\pi}_{+}{\rm
cos}(\psi)-\frac{1}{4}e^{-2i\phi}\hat{\pi}_{-}{\rm sin}(\psi){\rm
tan}(\psi) +\frac{3}{4}\hat{\pi}_{+}{\rm sin}(\psi){\rm tan}(\psi)+
\frac{1}{4}e^{-2i\phi}\hat{\pi}_{-}{\rm cos}(2\theta){\rm sin}(\psi){\rm tan}(\psi)\nonumber \\
&&+\frac{1}{4}\hat{\pi}_{+}{\rm cos}(2\theta){\rm sin}(\psi){\rm
tan}(\psi)- \frac{1}{2\sqrt{2}}e^{-i\phi}\hat{\pi}_{0}{\rm
sin}(2\theta){\rm sin}(\psi){\rm tan}(\psi)\\
A_{21}&=&\hat{\pi}_{-}{\rm cos}(\psi)+\hat{\pi}_{-}{\rm
cos}^2(\theta){\rm sin}(\psi){\rm tan}(\psi)-
\frac{1}{\sqrt{2}}e^{i\phi}\hat{\pi}_0 {\rm cos}(\theta){\rm sin}(\theta){\rm sin}(\psi){\rm tan}(\psi) \nonumber \\
&&+\frac{1}{2}\hat{\pi}_{-}{\rm sin}^2(\theta){\rm sin}(\psi){\rm
tan}(\psi)-\frac{1}{2}e^{2i\phi}\hat{\pi}_{+}{\rm sin}^2(\theta){\rm
sin}(\psi){\rm tan}(\psi) \\
A_{22}&=&-\frac{1}{\sqrt{2}}\hat{\pi}_0
{\rm cos}(\psi)+\frac{i}{\sqrt{2}}\hat{\pi}_0{\rm cos}(\theta){\rm
sin}(\psi) +\frac{i}{2}e^{-i\phi}\hat{\pi}_{-}{\rm sin}(\theta){\rm
sin}(\psi)+
\frac{i}{2}e^{i\phi}\hat{\pi}_{+}{\rm sin}(\theta){\rm sin}(\psi)\nonumber \\
&&+ \frac{i}{2}e^{-i\phi}\hat{\pi}_{-}{\rm cos}(\theta){\rm sin}(\theta){\rm sin}(\psi){\rm tan}(\psi)+
\frac{i}{2}e^{i\phi}\hat{\pi}_{+}{\rm cos}(\theta){\rm sin}(\theta){\rm sin}(\psi){\rm tan}(\psi)\nonumber \\
&&-\frac{1}{\sqrt{2}}\hat{\pi}_0 {\rm sin}^2(\theta){\rm
sin}(\psi){\rm tan}(\psi)
\end{eqnarray}

\bibliography{epsilonpaper7}

\end{document}